\documentclass[aps,prl,twocolumn,showpacs]{revtex4-2}

\usepackage{color}

\usepackage{graphicx,amsmath,amssymb}

\begin{document}

\title{Experimental noiseless quantum amplification of coherent states of light by two-photon addition and subtraction}

\author{ Michal Neset}
\affiliation{Department of Optics, Faculty of Science, Palack\'y University, 17.\ listopadu 12, 77900  Olomouc, Czech Republic}

\author{Jiří Fadrný}
\affiliation{Department of Optics, Faculty of Science, Palack\'y University, 17.\ listopadu 12, 77900  Olomouc, Czech Republic}

\author{Martin Bielak}
\affiliation{Department of Optics, Faculty of Science, Palack\'y University, 17.\ listopadu 12, 77900  Olomouc, Czech Republic}

\author{Jaromír Fiurášek}
\affiliation{Department of Optics, Faculty of Science, Palack\'y University, 17.\ listopadu 12, 77900  Olomouc, Czech Republic}

\author{Miroslav Ježek}
\affiliation{Department of Optics, Faculty of Science, Palack\'y University, 17.\ listopadu 12, 77900  Olomouc, Czech Republic}

\author{Jan Bílek}
\affiliation{Department of Optics, Faculty of Science, Palack\'y University, 17.\ listopadu 12, 77900  Olomouc, Czech Republic}

\begin{abstract}

Noiseless quantum amplifiers are probabilistic quantum devices that enhance amplitude of coherent states without adding any noise, which has far reaching applications in quantum optics and quantum information processing. Here, we report on experimental implementation of an  advanced noiseless quantum amplifier for coherent states of light that is  based on conditional addition of two photons followed by conditional subtraction of two photons. We comprehensively characterize the noiselessly amplified coherent states via quantum state tomography and analyze the amplification gain and noise properties of the amplifier. We observe very good agreement between the experiment and theoretical predictions. 
Our work reveals  that sequences of multiple photon additions and subtractions represent an efficient and experimentally feasible alternative to multiplexing that was originally proposed to boost the performance of noiseless quantum amplifiers.  Beyond noiseless quantum amplification, our experiment represents a significant step forward towards engineering complex quantum operations on traveling light beams by coherent combinations of various sequences of multiphoton additions and subtractions.
\end{abstract}

\maketitle

The laws of quantum physics imply that phase insensitive amplification of coherent optical signals is unavoidably accompanied by addition of thermal noise. Specifically, a linear quantum-limited amplifier with amplitude gain $g$ adds on average $\bar{n}=g^2-1$ thermal photons to the amplified mode.
This noisy nature of ordinary optical amplifiers  usually precludes their utilization in quantum communication and information processing. In 2009, Tim Ralph and Austin Lund put forward the concept of noiseless quantum amplifier \cite{Ralph2009}, an intriguing  device that amplifies the amplitude of a coherent state $|\alpha\rangle$ without adding any noise. While deterministic noiseless amplifiers do not exist, it is possible to design probabilistic noiseless amplifiers that approximately implement the operation $|\alpha\rangle \rightarrow |g\alpha\rangle$
for some subset of the coherent states \cite{Ferreyrol2010,Xiang2010,Zavatta2010,Usuga2010,Osorio2012,Kocsis2013,Chrzanowski2014,Haw2016,Fadrny2024}. Besides being of fundamental interest, noiseless quantum amplifiers find diverse applications in optical quantum information processing. Examples include improved performance of quantum key distribution \cite{Gisin2010,Blandino2012,Xu2013}, suppression of losses in optical quantum communication \cite{Ralph2011,Micuda2012,Zhang2012,Seshadreesan2020}, distillation of  entanglement \cite{Xiang2010,Ulanov2015,Seshadreesan2019,Guanzon2023}, and probabilistic cloning of coherent states \cite{Haw2016}.

The ideal probabilistic noiseless quantum amplifier is described by a non-unitary quantum operation $\hat{G}=g^{\hat{n}}$ which is diagonal in Fock basis, and $\hat{n}$ denotes the photon number operator. The experimentally realized noiseless amplifiers then approximate this still unphysical operation by performing suitable modulations of Fock-state amplitudes of the input state. Two main  approaches are pursued. The first one is based on the generalized quantum scissors scheme \cite{Pegg1998}, where ancilla single-photon or multiphoton Fock states drive the noiseless amplification via quantum interference and photon detection \cite{Ferreyrol2010,Xiang2010,Osorio2012,Kocsis2013,Guanzon2022}. 
The second main approach is based on combination of conditional photon addition \cite{Zavatta2004,Barbieri2010,Kumar20013} and photon subtraction \cite{Wenger2004,Ourjoumtsev2006,Nielsen2006,Wakui2007}, which allows to engineer a high-fidelity noiseless quantum amplifier with superior performance \cite{Fiurasek2009,Marek2010,Zavatta2010}. Experimental demonstrations of this second class of noiseless quantum amplifiers include an amplifier based on sequence of single-photon addition and single-photon subtraction \cite{Zavatta2010}, and an amplifier based on random coherent displacements combined with subtraction of multiple photons \cite{Usuga2010}. Very recently, we have experimentally implemented conditional addition of up to three photons into a propagating mode of optical field \cite{Fadrny2024} and we have verified that sole multi-photon addition can serve as an approximate noiseless quantum amplifier for coherent states with not too small amplitude \cite{Park2016}.

In the present work we further significantly push forward the experimental capabilities of conditional photon addition and subtraction. Namely, we experimentally demonstrate noiseless quantum amplifier based on sequence of conditional addition of two photons followed by conditional subtraction of two photons. Remarkably, we find that, assuming the same total non-Gaussian resources, our noiseless amplifier exhibits essentially the same performance as a much more complex interferometric scheme, where the signal is split into two modes, each mode is noiselessly amplified, and the two modes are interferometrically recombined. Combination of multiple photon additions and subtractions therefore represents an efficient and experimentally feasible alternative to multiplexing that was originally proposed to boost the performance of noiseless quantum amplifiers \cite{Ralph2009,Xiang2010}. Our work not only provides a new noiseless quantum amplifier for quantum states of freely propagating optical modes, but it paves the way towards realization of complex quantum operations based on coherent superpositions  of various sequences of multiple photon additions and subtractions \cite{Fiurasek2009,Zavatta2009,Costanzo2017,Biagi2022}.

\begin{figure*}[t!]
  \centering\includegraphics[width=\linewidth]{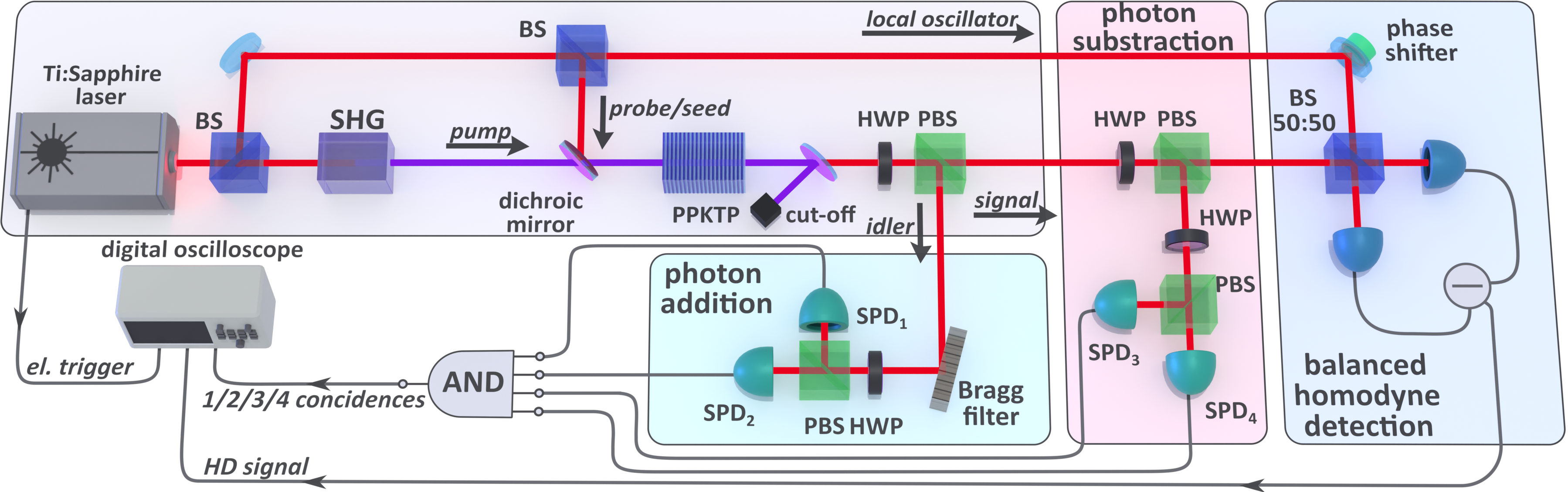}
  \caption{Experimental setup. BS, beam splitter; SHG, second-harmonic generation; PPKTP, periodically polled potassium titanyl phosphate crystal, HWP, half-wave plate; PBS, polarizing beam splitter; SPD, single-photon detector; AND, coincidence logic gate.}
  \label{figexpsetup}
\end{figure*}

Quantum states generated in our experiment  are prepared from input coherent states $|\alpha\rangle$ by conditional addition of $m$ photons followed by conditional subtraction of $m$ photons. These states can be expressed as 
\begin{equation}
|\psi_m(\alpha)\rangle= \frac{1}{\sqrt{M_m(\alpha)}}\hat{a}^{m}\hat{a}^{\dagger m}|\alpha\rangle,
\label{psim}
\end{equation}
where $M_m(\alpha)=\langle \alpha|(\hat{a}^m\hat{a}^{\dagger m})^2|\alpha\rangle$ is a normalization factor.
The operator $\hat{G}_m= \hat{a}^{m}\hat{a}^{\dagger m}$ is diagonal in Fock basis and can be expressed in terms of photon number operator $\hat{n}=\hat{a}^\dagger \hat{a}$, $\hat{G}_m=\prod_{j=1}^m(\hat{n}+j)$. In our experiment, we focus on the addition and subtraction of two photons ($m=2$). For comparison, we also implement the sequence of single-photon addition followed by single-photon subtraction ($m=1$), which was first experimentally realized by Zavatta \emph{et al.} \cite{Zavatta2010}. The corresponding operators $\hat{G}_1$ and $\hat{G}_2$ read
\begin{equation}
\hat{G}_1=\hat{n}+1, \qquad 
\hat{G}_2=\hat{n}^2+3\hat{n}+2.
\end{equation}

Our experimental setup is depicted in Fig.~\ref{figexpsetup}. The primary light source is provided by pulsed Ti:Sapphire laser Mira (Coherent) that emits 1.5 ps pulses at a central wavelength of 800 nm with repetition rate of 76~MHz. Most of the laser power is converted to 400 nm in a second harmonic generation module and pumps periodically poled nonlinear crystal PPKTP, where correlated photon pairs are generated in the process of collinear type-II parametric down conversion. The pump light is spectrally filtered out at the crystal output and the signal and idler modes are spatially separated at a polarizing beam splitter. A small part of the laser light at 800 nm is injected into the nonlinear crystal to  prepare the input signal mode in coherent state $|\alpha\rangle$.  Conditional addition of $m$ photons is heralded by detection of $m$ photons in the output idler mode. To subtract $m$ photons, a small part of the signal is tapped off by an unbalanced beam splitter formed by a half-wave plate and a polarizing beam splitter. Subtraction of $m$ photons is heralded by detection of $m$ photons in the auxiliary output mode of this beam splitter. 
We detect the heralding photons with  single-photon avalanche diodes that can only distinguish the presence and absence of photons. We therefore employ spatial multiplexing and split each detected mode onto two detectors to enable heralding on detection of two photons. 
We adjust the intensity transmittance $T$  of the photon-subtracting beam splitter such as to achieve high generation rate while avoiding saturation of the heralding detectors.
 In the experiment, the transmittance $T$  thus slightly varies and ranges from $0.9$ to $0.95$ which is sufficiently large to  keep the unwanted contributions from subtractions of more than $m$ photons acceptably small. 

The output signal mode is fed to a home-built balanced homodyne detector with a 12~dB signal-to-noise ratio and 100~MHz bandwidth and the generated state is comprehensively characterized by homodyne tomography. We utilize maximum likelihood reconstruction algorithm to determine the generated state $\hat{\rho}$ from the sampled quadrature statistics \cite{Jezek2003,Hradil2004,Lvovsky2004}. We also reconstruct the input coherent states from quadrature samples recorded when the heralding detectors do not click. We estimate that the total efficiency of the homodyne detection is $57\%$   \cite{Fadrny2024} and compensate the detection losses during the quantum state reconstruction. For additional experimental details, see \cite{Fadrny2024}.

Due to the non-unit transmittance of the photon-subtraction beam splitter, the conditionally generated states (\ref{psim}) are affected by additional noiseless attenuation $t^{\hat{n}}$, where $t=\sqrt{T}$ is the amplitude transmittance \cite{Fiurasek2009,Micuda2012,Nunn2021}. This attenuation only reduces the effective amplitude of the seed coherent state to $t\alpha$. 
Since we reconstruct the seed coherent states from measurements at the setup output, we in fact observe the attenuated seed coherent state $|t\alpha\rangle$. This means that the effect of (noiseless) attenuation is included in our determination of the effective input complex amplitude and we can therefore consider the ideal operations $\hat{G}_m$ in the subsequent analysis. 

\begin{figure}
\centerline{\includegraphics[width=\linewidth]{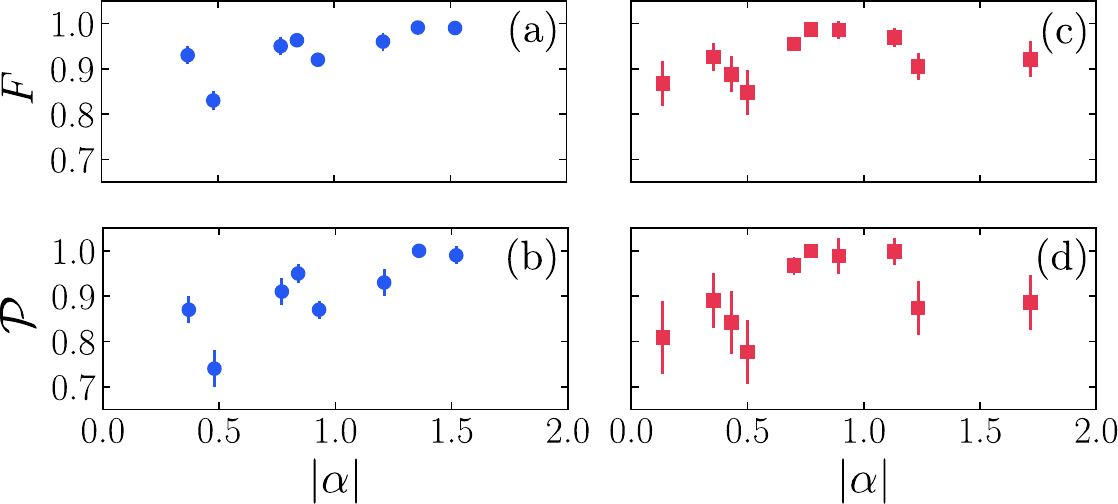}}
\caption{Fidelities $F$ and purities $\mathcal{P}$ of states generated from input coherent states with amplitude $\alpha$ by  sequence of single photon addition and subtracition (a,b) or by sequence of two-photon addition and two-photon subtraction (c,d). }
\label{figFP}
\end{figure}

\begin{figure}[!t!]
  \centering\includegraphics[width=\linewidth]{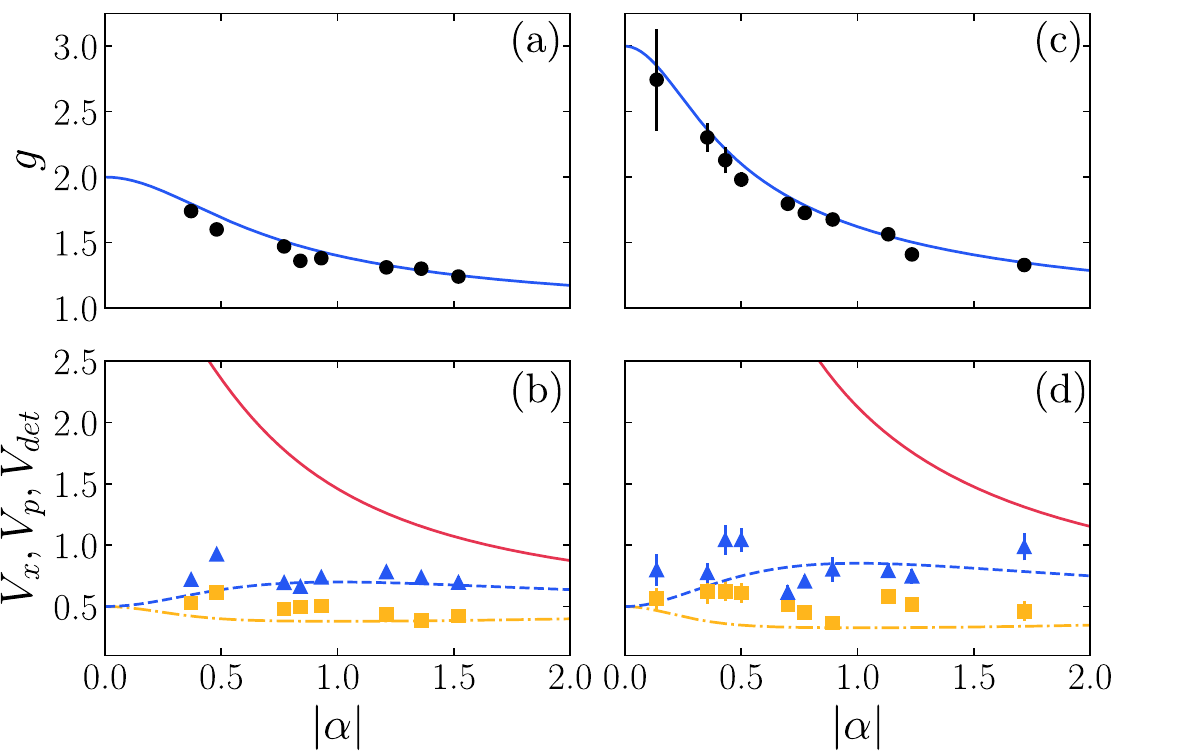}
\caption{Amplification gains $g_m$ and amplitude and phase quadrature variances $V_x$ (yellow squares) and $V_p$ (blue triangles) of the noiselessly amplified coherent states $\hat{a}^m \hat{a}^{\dagger m}|\alpha\rangle$ are plotted for $m=1$ (a,b) and $m=2$ (c,d). Lines indicate theoretical predictions, symbols represent experimental data. The solid red lines in panels (b,d), indicate the quadrature variance $V_{\mathrm{det}}$ achievable by deterministic phase-insensitive amplifier with gain $g_m$ .}
\label{figgainsVxp}
\end{figure}

In our experiment, we have generated states $|\psi_2(\alpha)\rangle$ and $|\psi_1(\alpha)\rangle$ for input coherent states with various amplitudes $\alpha$ ranging from $0.14$ to $1.72$.  We consistently observe very high fidelities $F_m=\langle \psi_m(\alpha)|\hat{\rho}|\psi_m(\alpha)\rangle$ and purities$\mathcal{P}=\mathrm{Tr}[\hat{\rho}^2]$  of the generated states , see Fig.~\ref{figFP}. The average state fidelity and purity of states obtained by addition and subtraction of two photons  reads $\bar{F}_2=0.93\pm 0.01$, $\bar{\mathcal{P}}_2=0.90\pm 0.02$, while for the states generated by combination of single-photon  addition and subtraction we get $\bar{F}_1=0.94\pm 0.01$,  $\bar{\mathcal{P}}_1=0.91\pm 0.01$. Typical success probability of implementation of the sequence $\hat{a}\hat{a}^\dagger$ is $10^{-5}$, an order of magnitude improvement compared to Ref. \cite{Zavatta2010}. For the sequence of two photon additions and subtractions typical success probability drops to $10^{-9}$. Nevertheless, thanks to the high stability of our setup, we can perform measurements on time scale of hours and collect enough quadrature samples to characterize the generated states. 

The operations $\hat{G}_m$ act as approximate noiseless amplifiers that amplify the amplitude of the coherent state $|\alpha\rangle$ while preserving its purity.  The  amplification gain of the amplifier is defined as the ratio of complex amplitudes of output and input states,
\begin{equation}
g_m(\alpha)= \frac{\langle \psi_m(\alpha)| \hat{a}|\psi_m(\alpha)\rangle}{\alpha}.
\end{equation}
 Explicit calculation yields  \cite{Fiurasek2009,Zavatta2010}
\begin{eqnarray}
& \displaystyle g_1(\alpha)=1+\frac{1+|\alpha|^2}{1+3|\alpha|^2+|\alpha|^4},& \nonumber \\[2mm]
&\displaystyle g_2(\alpha)=1+\frac{2(2+6|\alpha|^2+|\alpha|^4)(|\alpha|^2+2)}{4+32|\alpha|^2+38|\alpha|^4+12|\alpha|^6+|\alpha|^8}.&
\end{eqnarray}
The amplification gains $g_1$ and $g_2$ are monotonically decreasing functions of $|\alpha|$ and they approach $1$ in the limit of large $|\alpha|$. Highest gain is achieved in the opposite limit $|\alpha|\rightarrow 0$, when $g_2(0)=3$ while $g_1(0)=2$. The experimentally determined amplification gains are plotted in Fig.~\ref{figgainsVxp}(a,c). The experimental results are in very good agreement with theory and for $m=1$ we largely reproduce the results obtained previously by Zavatta \emph{et al.} \cite{Zavatta2010}. Note that with the sequence $\hat{a}^2\hat{a}^{\dagger 2}$ we achieve experimental gains significantly higher than $2$, which is impossible to achieve by the sequence $\hat{a}\hat{a}^\dagger$. Due to noiseless attenuation, the true amplification gains are smaller by a factor of $\sqrt{T}$ with respect to the results plotted in Fig. ~3. Since $T$ varies in the experiment, the reduction factor varies from $0.95$ to $0.97$.

\begin{figure}[!t!]
\includegraphics[width=\linewidth]{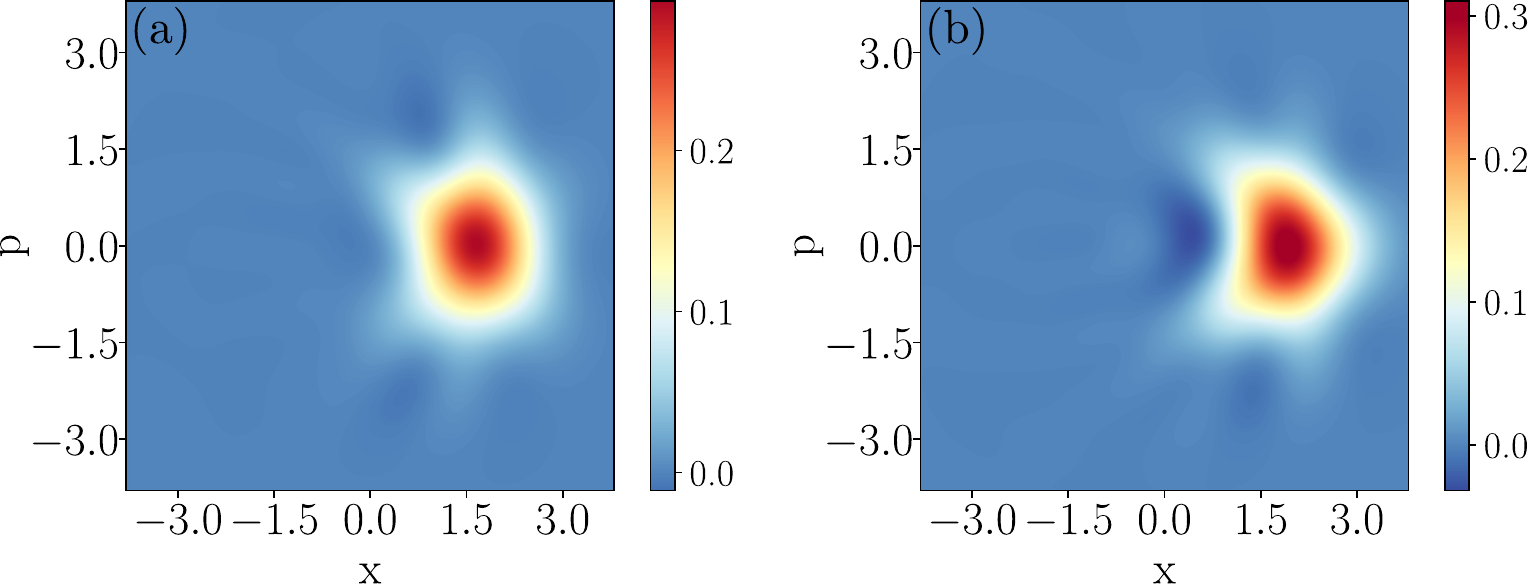}
\caption{Experimental Wigner functions of noiselessly amplified coherent state with input amplitude $\alpha=0.77$. Results are shown for amplification by operations  $\hat{a} \hat{a}^\dagger$ (a) and $\hat{a}^2 \hat{a}^{\dagger 2}$ (b). }
\label{figtwoWigner}
\end{figure}

Besides the amplification gain, the amplifier is also characterized by the variances $V_x$ and $V_p$ of amplitude and phase quadratures of the amplified state $\hat{x}=\frac{1}{\sqrt{2}}(\hat{a}e^{-i\theta}+\hat{a}^\dagger e^{i\theta})$ and $\hat{p}=\frac{i}{\sqrt{2}}(\hat{a}^\dagger e^{i\theta}-\hat{a}e^{-i\theta})$, where $\theta=\arg(\alpha)$. The variances are normalized such that $V_x=V_p=\frac{1}{2}$ for vacuum or coherent state. Explicit analytical formulas for the quadrature variances of the states  $|\psi_1(\alpha)\rangle$ and $|\psi_2(\alpha)\rangle$ are provided in the Appendix.
Importantly, the quadrature variances of noiselessly amplified states are much smaller than the variance achievable by the optimal deterministic linear quantum amplifier with the same gain, $V_{\mathrm{det}}=g_m^2-\frac{1}{2}$ .
The superior noise performance of our noiseless quantum amplifiers is illustrated in Fig.~\ref{figgainsVxp}(b,d).

Since the states $|\psi_m(\alpha)\rangle$ are not exactly Gaussian, the generated states are not minimum uncertainty states and $V_xV_p>\frac{1}{4}$ in general. Theory predicts that  $V_p>V_x$ and  $V_x<\frac{1}{2}$  for both amplifiers, i.e. the states $|\psi_m(\alpha)\rangle$ are slightly quadrature squeezed in the amplitude quadrature \cite{Zavatta2004,Fadrny2024}. Our experimental observations are consistent with these predictions. Examples of Wigner functions of the conditionally generated noiselessly amplified coherent states are provided in Fig.~\ref{figtwoWigner} and they exhibit the expected elliptical shape. The noiselessly amplified states are overall rather similar to Gaussian states, which can be quantified by their fidelity with a coherent state with the same complex amplitude. For the experimentally generated states we find that their coherent-state fidelities typically exceed $90\%$ for both studied amplifiers. Theory predicts $F_{\mathrm{coh}}>0.981 $ for $ m=1$ and
$F_{\mathrm{coh}}>0.963$ for $m=2$ for all $\alpha$. This further confirms that the combinations of photon additions and subtractions serve as high-quality noiseless quantum amplifiers for coherent states of light.

\begin{figure}[!t!]
  \centerline{\includegraphics[width=\linewidth]{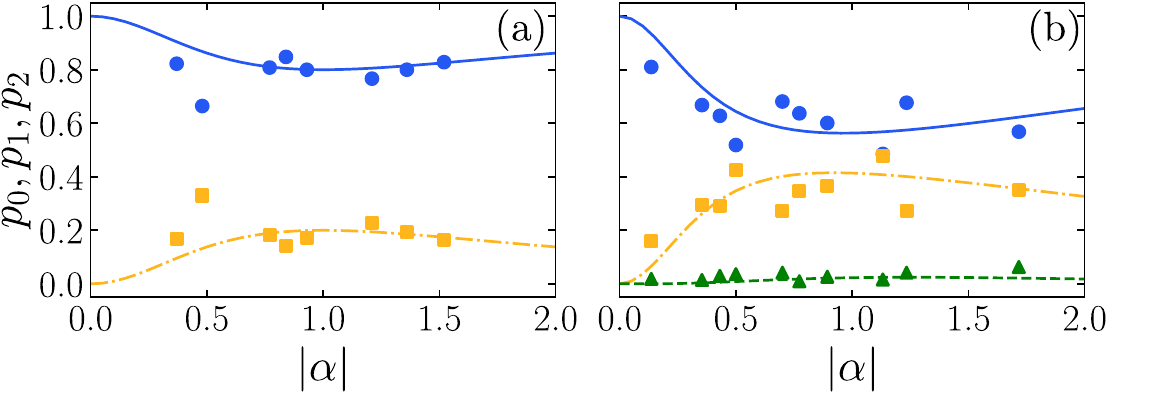}}
\caption{Photon number distribution $\tilde{p}_n$ of the inversely displaced noiselessly amplified coherent states $\hat{D}(-\alpha)\hat{G}_1|\alpha\rangle$ (a) and  $\hat{D}(-\alpha)\hat{G}_2|\alpha\rangle$ (b). Lines indicate theoretical predictions and symbols show experimental data. Results are plotted for $n=0$ (blue solid line), $n=1$ (yellow dot-dashed  line), and $n=2$ (green dashed line). }
\label{figtwopn}
\end{figure}

The noiselessly amplified coherent state $\hat{G}_2|\alpha\rangle$ can be equivalently expressed as a coherently displaced superposition of vacuum, single-photon, and two-photon states,
\begin{eqnarray}
\hat{a}^2\hat{a}^{\dagger 2}|\alpha\rangle&=& \hat{D}(\alpha) \left[ (2+4|\alpha|^2+|\alpha|^4)|0\rangle+2\alpha(2+|\alpha|^2)|1\rangle \right.\nonumber \\
& &\left.+\sqrt{2}\alpha^2|2\rangle\right].
\label{psi2}
\end{eqnarray}
Here $\hat{D}(\alpha)=e^{\alpha \hat{a}^\dagger-\alpha^\ast \hat{a}}$ denotes the coherent displacement operator. Similarly, the state $\hat{G}_1|\alpha\rangle$ can be written as
\begin{equation}
\hat{a}\hat{a}^\dagger|\alpha\rangle=\hat{D}(\alpha)[(1+|\alpha|^2)|\alpha\rangle+\alpha|1\rangle].
\end{equation}
In Fig.~\ref{figtwopn} we plot the normalized probabilities $\tilde{p}_n $ of Fock states $|n\rangle$ in the finite superposition state $\hat{D}(-\alpha)|\psi_m(\alpha)\rangle$. The vacuum probability $p_0$ is dominant for both amplifiers, however with two additions and subtractions we get closer to balanced contribution of vacuum and single-photon components. The two-photon contribution $p_2$ is vanishing in theory for $m=1$ and is very small for $m=2$, with maximum $p_{2,\mathrm{max}}\approx 0.025$. Nevertheless, the two-photon term in Eq. (\ref{psi2}) is important as it for instance enables to achieve amplification gain higher than $1.5$ at $|\alpha|=1$. The experimentally generated states are well concentrated into the subspace spanned by the first $m+1$ Fock states after the inverse displacement $\hat{D}(-\alpha)$. We typically observe $p_0+p_1>0.99$ for $m=1$ and $p_0+p_1+p_2 >0.98$ for $m=2$.

\begin{figure}
  \centering\includegraphics[width=0.7\linewidth]{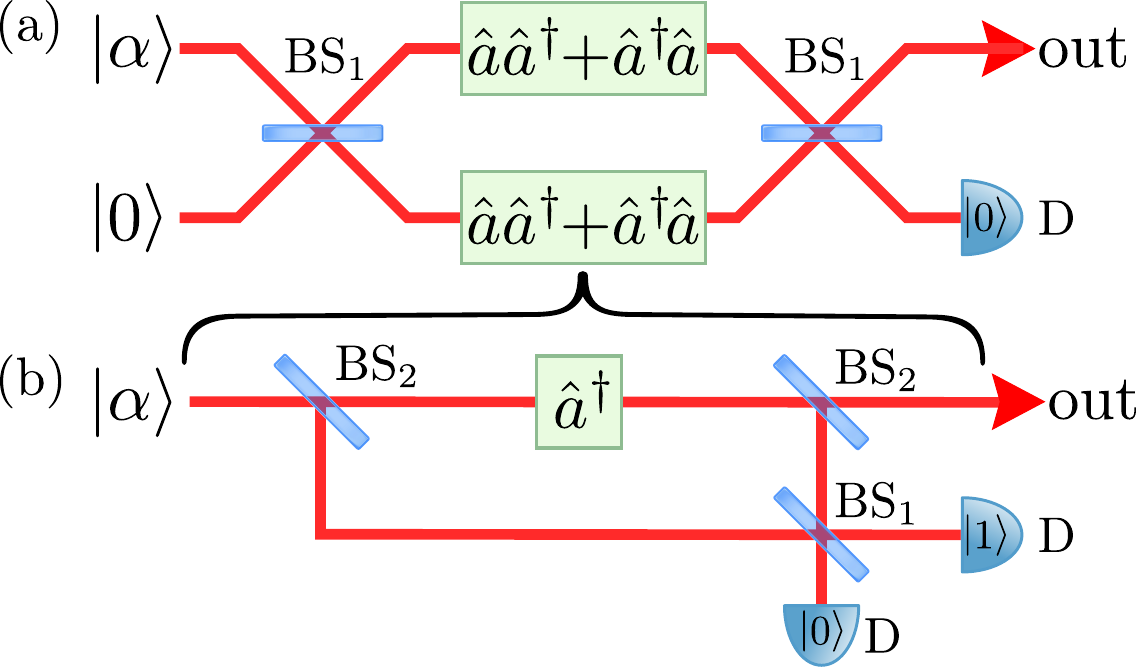}
\caption{Noiseless quantum amplifier based on multiplexing (a) and scheme for implementation of the coherent superposition  $\hat{a}\hat{a}^\dagger+\hat{a}^\dagger\hat{a}$ (b). The schemes include balanced beam splitters BS$_1$, unbalanced beam splitters BS$_2$,  and single-photon detectors D.}
\label{figmultiplex}
\end{figure}

To further improve the performance of the noiseless quantum amplifiers, it was proposed in Refs. \cite{Ralph2009,Xiang2010} to first split the amplified coherent signal into $N$ modes, noiselessly amplify each mode and then recombine the state back into a single mode by an array of $N-1$ beam splitters followed by projections of all auxiliary output modes onto vacuum.   It is instructive to compare the performance of our noiseless amplifier $\hat{G}_2=\hat{a}^2\hat{a}^{\dagger 2}$ with the scheme that utilizes multiplexing into two modes, as illustrated in Fig.~\ref{figmultiplex}. Each mode inside the interferometer in Fig.~\ref{figmultiplex}(a) is approximately noiselessly amplified by the operation $\hat{G}_1^\prime=\hat{a}\hat{a}^\dagger+\hat{a}^\dagger\hat{a}=2\hat{n}+1$ which involves one  photon addition and one photon subtraction. The amplifier $\hat{G}_1^\prime$ achieves the same nominal gain $g=3$ for small $|\alpha|$ as the amplifier $\hat{G}_2$ which is important  for fair comparison of both approaches. Note that the operation $\hat{G}_1^\prime$ has been implemented experimentally in the context of experimental tests of quantum commutation and anti-commutation relations \cite{Zavatta2009}. As illustrated in Fig.~\ref{figmultiplex}(b), implementation of the operation $\hat{G}_1^\prime$ requires an interferometric scheme, where photon subtraction is attempted both before and after the photon addition, and the which way information about the subtracted photon is subsequently erased by interference at a beam splitter.  Since the setup in Fig.~\ref{figmultiplex} contains two copies of the operation $\hat{G}_1^\prime$  that must succeed simultaneously, the setup involves two photon additions and two photon subtractions in total, and also one projection onto vacuum.

\begin{figure}
  \centering\includegraphics[width=\linewidth]{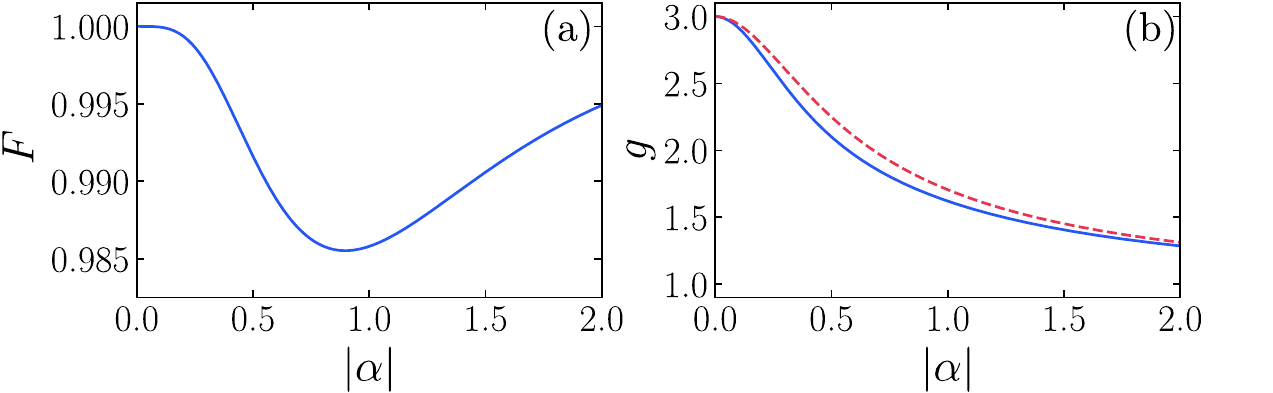}
\caption{Comparison of noiseless quantum amplifiers $\hat{G}_2$ and $\hat{G}_2^\prime$. Mutual fidelity of the noiselessly amplified coherent states produced by the two amplifiers (a),  and the amplification gains $g_2$ (blue solid line) and $g_2^\prime$ (red dashed line) (b) are plotted as functions of $|\alpha|$.}
\label{figcomparison}
\end{figure}

The effective operation on the input mode implemented by the multiplexed scheme in Fig.~\ref{figmultiplex} can be expressed as
\begin{equation}
\hat{G}_2^\prime=_B\!\!\langle 0| \hat{U}_{\mathrm{BS}}^\dagger  (2\hat{n}_A+1) (2\hat{n}_B+1)\hat{U}_{\mathrm{BS}} |0\rangle_B,
\end{equation}
where A and B label the two modes and $\hat{U}_{\mathrm{BS}}$ denotes the unitary operation realized by balanced beam splitter.  
After some algebra, we find that 
\begin{equation}
\hat{G}_2^\prime=\hat{n}^2+\hat{n}+1.
\label{G2prime}
\end{equation}
The state generated by this operation from a coherent state $|\alpha\rangle$ reads
\begin{equation}
\hat{G}_2^\prime|\alpha\rangle=\hat{D}(\alpha)\left[(1+|\alpha|^2)^2|0\rangle+2\alpha(1+|\alpha|^2)|1\rangle+\sqrt{2}\alpha^2|2\rangle\right].
\label{psi2prime}
\end{equation}
This state is very similar to the state (\ref{psi2}) produced in our experiment by two-photon additon followed by two-photon subtraction. This is illustrated in 
Fig.~\ref{figcomparison} where we plot the mutual fidelity of the states (\ref{psi2}) and (\ref{psi2prime}) and we also compare the amplification gains achieved by the two schemes. Based on these results we can conclude that our stable and robust scheme is fully equivalent to the much more complicated interferometric setup in Fig.~\ref{figmultiplex}.

In summary, we have successfully implemented advanced conditional quantum operation consisting of two-photon addition followed by two-photon subtraction and we have verified that this operation noiselessly amplifies coherent states. Our advanced noiseless amplifier exhibits high amplification gain, high fidelity and purity, and quadrature variances  of the noiselessly amplified coherent states remain close to the minimum uncertainty limit.  The success probability of the scheme can be further substantially increased with respect to our current proof-of-principle experiment by utilizing  spatially multiplexed superconducting detectors with detection efficiency exceeding $90\%$ and by employing a crystal that can withstand higher pump powers. 
Beyond noiseless amplification, our work represents important step towards flexible engineering of quantum operations via combinations of various sequences of multiphoton additions and subtractions \cite{Fiurasek2009,Costanzo2017}.

\begin{acknowledgments}
We acknowledge the financial support of the Czech Science Foundation (Project No. 21-23120S) and the Palacký University under Project No. IGA-PrF-2024-008. J. Fadrn\'{y} and J. Fiur\'{a}\v{s}ek
acknowledge the project 8C22002 (CVSTAR) of MEYS of the Czech Republic, which has received funding from the European Union’s Horizon 2020 Research and Innovation Programme under Grant Agreement no. 731473 and 101017733.
\end{acknowledgments}

\begin{widetext}

\bigskip 

\begin{center}{\textbf{END MATTER}}
\end{center}

\medskip

\emph{Appendix:} Here we provide analytical formulas for quadrature variances of the noiselessly amplified coherent states for the various noiseless amplifiers discussed in the main text. 
Let us first consider the amplifier $\hat{G}_1=\hat{n}+1$. Variances of amplitude and phase quadratures of the noiselessly amplified coherent state  $\hat{G}_1 |\alpha\rangle$ read
\begin{equation}
{V}_x=\frac{1}{2}-\frac{|\alpha|^2(1+|\alpha|^2+|\alpha|^4)}{(1+3|\alpha|^2+|\alpha|^4)^2}, \qquad {V}_p=\frac{1}{2}+\frac{|\alpha|^2}{1+3|\alpha|^2+|\alpha|^4}.
\end{equation}
Next we present results for the noiseless amplifier $\hat{G}_2=\hat{a}^2\hat{a}^{\dagger 2}=\hat{n}^2+3\hat{n}+2$. Variances of the amplitude and phase quadratures of the state $\hat{G}_2|\alpha\rangle$ can be expressed as follows,
\begin{eqnarray}
V_x&=& \frac{1}{2}-  \frac{2|\alpha|^2(24+72|\alpha|^2+200|\alpha|^4+180|\alpha|^6+76|\alpha|^8+14|\alpha|^{10}+|\alpha|^{12})}{(4+32|\alpha|^2+38|\alpha|^4+12|\alpha|^6+|\alpha|^8)^2}, \nonumber \\[2mm]
V_p&=&\frac{1}{2}+\frac{2|\alpha|^2(6+6|\alpha|^2+|\alpha|^4)}{4+32|\alpha|^2+38|\alpha|^4+12|\alpha|^6+|\alpha|^8}.
\end{eqnarray}
For completeness we provide also expressions for the noiseless amplifier $\hat{G}_2^\prime=\hat{n}^2+\hat{n}+1$  based on multiplexing. Variances of the amplitude and phase quadratures of the state $\hat{G}_2^\prime|\alpha\rangle$ are given by
\begin{eqnarray}
V_x^\prime&=&\frac{1}{2}-\frac{2|\alpha|^2 (1+8|\alpha|^2+25|\alpha|^4+32|\alpha|^6+25|\alpha|^8+8|\alpha|^{10}+|\alpha|^{12})}{(1+8|\alpha|^2+16|\alpha|^4+8|\alpha|^6+|\alpha|^8)^2}, \nonumber \\[2mm]
V_p^\prime&=&\frac{1}{2}+\frac{2|\alpha|^2(1+4|\alpha|^2+|\alpha|^4)}{1+8|\alpha|^2+16|\alpha|^4+8|\alpha|^6+|\alpha|^8}.
\end{eqnarray}
Mutual fidelity of the noiselessly amplified coherent states $\hat{G}_2|\alpha\rangle$ and $\hat{G}_2^\prime|\alpha\rangle$ reads
\begin{equation}
F=\frac{(2+16|\alpha|^2+25|\alpha|^4+10|\alpha|^6+|\alpha|^8)^2}{(1+8|\alpha|^2+16|\alpha|^4+8|\alpha|^6+|\alpha|^8)(4+32|\alpha|^2+38|\alpha|^4+12|\alpha|^6+|\alpha|^8)}.
\end{equation}
We also provide explicit formula for the amplification gain of the noiseless amplifier $\hat{G}_2^\prime$,
\begin{equation}
g_2^\prime(\alpha)=1+\frac{2(1+5|\alpha^2|^2+5|\alpha|^4+|\alpha|^6)}{1+8|\alpha^2|^2+16|\alpha|^4+8|\alpha|^6+|\alpha|^8}
\end{equation}

\end{widetext}

\end{document}